\documentclass{article}
\usepackage{dafne99pro}
\begin{document}
\newcommand{\nn}{\noindent}
\newcommand{\cs}{\mbox{$\clubsuit$}}
\newcommand{\nl}{\nonumber \\}
\newcommand{\hf}{\hfill}
\newcommand{\naive}{na$\ddot{\imath}$ve}
\newcommand {\oa} {\mbox{${\cal O}( \alpha)$}}
\newcommand {\ho} {\mbox{${\cal O}( \alpha^{2})$}}
\hyphenation{brems-strah-lung}
\def\ss{\footnotesize}
\def\SS{\footnotesize}
\def\sss{\scriptscriptstyle}
\def\barp{{\raise.35ex\hbox{${\sss (}$}}---{\raise.35ex\hbox{${\sss )}$}}}
\def\bdbarp{\hbox{$B_d$\kern-1.4em\raise1.4ex\hbox{\barp}}}
\def\bsbarp{\hbox{$B_s$\kern-1.4em\raise1.4ex\hbox{\barp}}}
\def\dbarp{\hbox{$D$\kern-1.1em\raise1.4ex\hbox{\barp}}}
\def\dcp{D^0_{\sss CP}}
\def\dbar{{\overline{D^0}}}
\def\ks{K_{\sss S}}
\newcommand{\xd}{x_d}
\newcommand{\xs}{x_s}
\newcommand{\bd}{B_d^0}
\newcommand{\bdb}{\overline{B_d^0}}
\newcommand{\bs}{B_s^0}
\newcommand{\bsbar}{\overline{B_s^0}}
\newcommand{\bu}{B_u^\pm}
\newcommand{\beq}{\begin{equation}}
\newcommand{\eeq}{\end{equation}}
\newcommand{\absvcb}{\vert V_{cb}\vert}
\newcommand{\absvub}{\vert V_{ub}\vert}
\newcommand{\absvtd}{\vert V_{td}\vert}
\newcommand{\absvts}{\vert V_{ts}\vert}
\newcommand{\abseps}{\vert\epsilon\vert}
\newcommand{\epsp}{\epsilon^\prime/\epsilon}
\newcommand{\fbb}{f^2_{B_d}\hat{B}_{B_d}}
\newcommand{\fbbs}{f^2_{B_s}\hat{B}_{B_s}}
\newcommand{\fbd}{f_{B_d}}
\newcommand{\fbs}{f_{B_s}}
\newcommand{\fds}{f_{D_s}}
\def\rly#1{\mathrel{\raise.3ex\hbox{$#1$\kern-.75em\lower1ex\hbox{$\sim$}}}}
\def\lsim{\rly<}

\def \zpc#1#2#3{{\rm Z.~Phys.} {\bf C#1} (19#2) #3}
\def \plb#1#2#3{{\rm Phys.~Lett.} {\bf B#1} (19#2) #3}
\def \ibj#1#2#3{~#1, (19#2) #3}
\def \prl#1#2#3{{\rm Phys.~Rev.~Lett.} {\bf #1} (19#2) #3}
\def \prd#1#2#3{{\rm Phys.~Rev.} {\bf D#1} (19#2) #3} 
\def \npb#1#2#3{{\rm Nucl.~Phys.} {\bf B#1} (19#2) #3} 
\def\ijmp#1#2#3{{\rm Int.\ J.\ Mod.\ Phys.} {\bf A#1} (19#2) #3}
\def \stone{{\it B Decays}, edited by S. Stone (World Scientific, Singapore,
1994)}

\newread\epsffilein 
\newif\ifepsffileok 
\newif\ifepsfbbfound 
\newif\ifepsfverbose 
\newdimen\epsfxsize 
\newdimen\epsfysize 
\newdimen\epsftsize 
\newdimen\epsfrsize 
\newdimen\epsftmp 
\newdimen\pspoints 
\pspoints=1bp 
\epsfxsize=0pt 
\epsfysize=0pt 
\def\epsfbox#1{\global\def\epsfllx{72}\global\def\epsflly{72}%
 \global\def\epsfurx{540}\global\def\epsfury{720}%
 \def\lbracket{[}\def\testit{#1}\ifx\testit\lbracket
 \let\next=\epsfgetlitbb\else\let\next=\epsfnormal\fi\next{#1}}%
\def\epsfgetlitbb#1#2 #3 #4 #5]#6{\epsfgrab #2 #3 #4 #5 .\\%
 \epsfsetgraph{#6}}%
\def\epsfnormal#1{\epsfgetbb{#1}\epsfsetgraph{#1}}%
\def\epsfgetbb#1{%
%
%
\openin\epsffilein=#1
\ifeof\epsffilein\errmessage{I couldn't open #1, will ignore it}\else
%
%
 {\epsffileoktrue \chardef\other=12
 \def\do##1{\catcode`##1=\other}\dospecials \catcode`\ =10
 \loop
 \read\epsffilein to \epsffileline
 \ifeof\epsffilein\epsffileokfalse\else
%
%
 \expandafter\epsfaux\epsffileline:. \\%
 \fi
 \ifepsffileok\repeat
 \ifepsfbbfound\else
 \ifepsfverbose\message{No bounding box comment in #1; using defaults}\fi\fi
 }\closein\epsffilein\fi}%
%
%
\def\epsfclipstring{}
\def\epsfclipon{\def\epsfclipstring{ clip}}%
\def\epsfclipoff{\def\epsfclipstring{}}%
\def\epsfsetgraph#1{%
 \epsfrsize=\epsfury\pspoints
 \advance\epsfrsize by-\epsflly\pspoints
 \epsftsize=\epsfurx\pspoints
 \advance\epsftsize by-\epsfllx\pspoints
%
%
 \epsfxsize\epsfsize\epsftsize\epsfrsize
 \ifnum\epsfxsize=0 \ifnum\epsfysize=0
 \epsfxsize=\epsftsize \epsfysize=\epsfrsize
 \epsfrsize=0pt
%
%
 \else\epsftmp=\epsftsize \divide\epsftmp\epsfrsize
 \epsfxsize=\epsfysize \multiply\epsfxsize\epsftmp
 \multiply\epsftmp\epsfrsize \advance\epsftsize-\epsftmp
 \epsftmp=\epsfysize
 \loop \advance\epsftsize\epsftsize \divide\epsftmp 2
 \ifnum\epsftmp>0
 \ifnum\epsftsize<\epsfrsize\else
 \advance\epsftsize-\epsfrsize \advance\epsfxsize\epsftmp \fi
 \repeat
 \epsfrsize=0pt
 \fi
 \else \ifnum\epsfysize=0
 \epsftmp=\epsfrsize \divide\epsftmp\epsftsize
 \epsfysize=\epsfxsize \multiply\epsfysize\epsftmp
 \multiply\epsftmp\epsftsize \advance\epsfrsize-\epsftmp
 \epsftmp=\epsfxsize
 \loop \advance\epsfrsize\epsfrsize \divide\epsftmp 2
 \ifnum\epsftmp>0
 \ifnum\epsfrsize<\epsftsize\else
 \advance\epsfrsize-\epsftsize \advance\epsfysize\epsftmp \fi
 \repeat
 \epsfrsize=0pt
 \else
 \epsfrsize=\epsfysize
 \fi
 \fi
%
%
 \ifepsfverbose\message{#1: width=\the\epsfxsize, height=\the\epsfysize}\fi
 \epsftmp=10\epsfxsize \divide\epsftmp\pspoints
 \vbox to\epsfysize{\vfil\hbox to\epsfxsize{%
 \ifnum\epsfrsize=0\relax
 \includegraphics{#1}%
 \else
 \epsfrsize=10\epsfysize \divide\epsfrsize\pspoints
 \includegraphics{#1}%
 \fi
 \hfil}}%
\global\epsfxsize=0pt\global\epsfysize=0pt}%
%
%
 {\catcode`\%=12 \global\let\epsfpercent=
%
%
\long\def\epsfaux#1#2:#3\\{\ifx#1\epsfpercent
 \def\testit{#2}\ifx\testit\epsfbblit
 \epsfgrab #3 . . . \\%
 \epsffileokfalse
 \global\epsfbbfoundtrue
 \fi\else\ifx#1\par\else\epsffileokfalse\fi\fi}%
%
%
\def\epsfempty{}%
\def\epsfgrab #1 #2 #3 #4 #5\\{%
\global\def\epsfllx{#1}\ifx\epsfllx\epsfempty
 \epsfgrab #2 #3 #4 #5 .\\\else
 \global\def\epsflly{#2}%
 \global\def\epsfurx{#3}\global\def\epsfury{#4}\fi}%
%
%
\def\epsfsize#1#2{\epsfxsize}
%
%
\let\epsffile=\epsfbox
\def\att{t \bar{t}}
\def\app{p \bar{p}}
\def\rts{\sqrt{s}}
\def\mt{m_t}
\def\mb{m_b}
\def\mc{m_c}
\newcommand{\bksgam}{\ $B \to K^*+ \gamma$}
\newcommand{\brogam}{\ $B \to \rho+ \gamma$}
\def\BDSl{B \to D^* \ell \nu_\ell}
\def\vdvp{v \cdot v^\prime}
\def\xiaoo{\xi_{A_1}(\vdvp =1 )}
\def\Vbc{V_{cb}}
\newcommand{\Tosc}{T_{osc}}
\newcommand{\sqrts}{\sqrt{s}}
\newcommand{\bg}{\beta \gamma}
\newcommand{\xds}{x_i}
\newcommand{\Ds}{D_s^\pm}
\newcommand{\bb}{B^0 B^0}
\newcommand{\barbar}{{\overline{B^0}}\thinspace{\overline{B^0}}}
\newcommand{\barb}{B^0 {\overline{B^0}}}
\newcommand{\bbar}{$B^0$--${\overline{B^0}}$}
\newcommand{\Deltat}{\Delta t}
\newcommand{\delt}{\delta t}
\newcommand{\delmd}{\Delta M_d}
\newcommand{\delms}{\Delta M_s}
\newcommand{\ps}{10^{-12} s}
\newcommand{\zbbar}{Z^0 \to b {\overline{b}}}
\newcommand{\eebbx}{$e^+ e^- \to B {\overline{B}} X$}
\newcommand{\pbpbbx}{$p{\overline{p}} \to B {\overline{B}} X$}
\newcommand{\kkbar}{$K^0$--${\overline{K^0}}$}
\newcommand{\bdbdbar}{$B_d^0$--${\overline{B_d^0}}$}
\newcommand{\bsbsbar}{$B_s^0$--${\overline{B_s^0}}$}
\newcommand{\as}{\mbox{$\alpha_{\displaystyle s}$}}
\newcommand{\aso}{\mbox{$O(\alpha_{\displaystyle s})$}}
\newcommand{\ass}{\mbox{$O(\alpha_{\displaystyle s}^2)$}}
\newcommand{\asq}{\mbox{$\alpha_{\displaystyle s}(Q^2)$}}
\newcommand{\ee}{\mbox{$e^+e^-$}}
\newcommand{\cc}{\mbox{$c {\overline{c}}$}}
\newcommand{\qq}{\mbox{$q {\overline{q}}$}}
\newcommand{\jp}{\mbox{$J/\Psi$}}
\newcommand{\lqc}{\Lambda_{QCD}}
\newcommand{\pmi}{{\not{p}}_{\perp}}
\newcommand{\set}{\sum E_{\perp}}
\newcommand{\ptr}{p_{\perp}}
\newcommand{\sww}{\sin^2{\theta_W}}
\newcommand{\sw}{\sin{\theta_W}}
\begin{flushright}
DESY 00-026 \\
UdeM-GPP-TH-00-68\\
hep-ph/0002167\\
February 2000\\
\end{flushright}

\vspace{0.5cm}
\centerline{\bf CP VIOLATION AND QUARK MIXING}
\vspace{1.0cm}

\begin{center}
Ahmed Ali        \\
Deutsches Elektronen Synchrotron DESY, Hamburg \\
Notkestra\ss e 85, D-22603 Hamburg
\vspace{0.5cm}

David London \\
Laboratoire Ren\'e J.-A. L\'evesque, Universit\'e de
Montr\'eal, \\
\em Montr\'eal, QC, Canada H3C 3J7

\vspace{6.0cm}
Invited Talk. To be published in the Proceedings of the 3rd Workshop on
Physics and Detectors for DA$\Phi$NE, Frascati, Italy, Nov. 16 -- 19,
1999
\end{center}
\thispagestyle{empty}
%
\title{CP VIOLATION AND QUARK MIXING}
\begin{center}
\author{
Ahmed Ali        \\
{\em Deutsches Elektronen Synchrotron DESY, Hamburg} \\
\\
David London \\
{\em Laboratoire Ren\'e J.-A. L\'evesque, Universit\'e de
Montr\'eal,} \\ 
{\em Montr\'eal, QC, Canada H3C 3J7}
}
\end{center}
\maketitle
\baselineskip=11.6pt
%
\begin{abstract}
Measurements of CP asymmetries in $B$ decays will soon be made at $B$
factories and hadron machines. In light of this, we review and update
the profile of the CKM unitarity triangle and the resulting CP
asymmetries in $B$ decays. This is done both in the standard model and
in several variants of the minimal supersymmetric standard model
(MSSM), which are characterized by a single phase in the quark flavour
mixing matrix. After imposing present constraints on the parameters of
these models, the predicted ranges of $\sin 2 \beta$ in the standard
model and in the MSSM are found to be similar. However, these theories
may be distinguished by future precise measurements of the other two
CP-violating phases $\alpha$ and $\gamma$.
\end{abstract}
\baselineskip=14pt
%

\section{Introduction}

Within the standard model (SM), CP violation is due to the presence of
a nonzero complex phase in the Cabibbo-Kobayashi-Maskawa (CKM) quark
mixing matrix $V$\cite{CKM}. A particularly useful parametrization of
the CKM matrix, due to Wolfenstein\cite{Wolfenstein}, follows from
the observation that the elements of this matrix exhibit a hierarchy
in terms of $\lambda$, the Cabibbo angle. In this parametrization the
CKM matrix can be written approximately as
\beq
V \simeq \left(\matrix{
 1-{1\over 2}\lambda^2 & \lambda
 & A\lambda^3 \left( \rho - i\eta \right) \cr
 -\lambda ( 1 + i A^2 \lambda^4 \eta )
& 1-{1\over 2}\lambda^2 & A\lambda^2 \cr
 A\lambda^3\left(1 - \rho - i \eta\right) & -A\lambda^2 & 1 \cr}\right)~.
\label{CKM}
\eeq
The allowed region in $\rho$--$\eta$ space can be elegantly displayed
using the so-called unitarity triangle (UT). The unitarity of the CKM
matrix leads to the following relation:
\beq
V_{ud} V_{ub}^* + V_{cd} V_{cb}^* + V_{td} V_{tb}^* = 0~.
\eeq
Using the form of the CKM matrix in Eq.~(\ref{CKM}), this can be
recast as
\beq
\label{trianglerel}
\frac{V_{ub}^*}{\lambda V_{cb}} + \frac{V_{td}}{\lambda V_{cb}} = 1~,
\eeq
which is a triangle relation in the complex plane (i.e.\
$\rho$--$\eta$ space). With the experimental precision expected in
future $B$ (and $K$) decays, it may become necessary to go beyond
leading order in $\lambda$ in the Wolfenstein parametrization given
above. To this end, we follow here the prescription of Buras et
al.\cite{BLO94}: defining $\bar{\rho} \equiv \rho(1-\lambda^2/2)$ and
$\bar{\eta} \equiv \eta(1-\lambda^2/2)$, we have
\beq
V_{us} = \lambda,~~~V_{cb}=A\lambda^2,~~~V_{ub}=A\lambda^3(\rho -i
\eta),~~~V_{td} = A\lambda^3(1- \bar{\rho} -i \bar{\eta}) 
\label{nlo-wolf}
\eeq
The key point here is that the matrix elements $V_{us}, V_{cb}$ and
$V_{ub}$ remain unchanged, but $V_{td}$ is renormalized in going from
leading order (LO) to next-to-leading order (NLO). The apex of the UT
is now defined by $(\bar{\rho},\bar{\eta})$.

Constraints on $\bar{\rho}$ and $\bar{\eta}$ come from a variety of
sources. For example, $|V_{cb}|$ and $|V_{ub}|$ can be extracted from
semileptonic $B$ decays, and $|V_{td}|$ is at present probed in
\bdbdbar\ mixing. The interior CP-violating angles $\alpha$, $\beta$
and $\gamma$ can be measured through CP asymmetries in $B$
decays. Additional constraints come from CP violation in the kaon
system ($\abseps$), as well as \bsbsbar\ mixing.

A profile of the unitarity triangle was presented by us in early
1999\cite{AL99}. This analysis was done at NLO precision, taking into
account the state-of-the-art calculations of the hadronic matrix
elements from lattice QCD and data available at that time.
Subsequently, an improved lower limit $\delms > 14.3~{\rm (ps)}^{-1}$
was reported at the Lepton-Photon symposium in the summer of
1999\cite{Blaylock99}. In this report, we update the results of our
1999 CKM-unitarity fits by incorporating this new limit on
$\delms$. As we shall see here, this measurement tightens the
constraints on the CKM parameters. The other new ingredient in our
fits is that we now use the improved Wolfenstein parametrization given
in Eq.~(\ref{nlo-wolf}). We also compare our results with two other
recent fits in which the new $\delms$-limit has been
incorporated\cite{PS99,herab-012}, but which differ from us in details
which we shall specify below.

If new physics (of any type) is present, the principal way in which it
can enter in flavour physics is via new contributions, possibly with
new phases, to \kkbar, \bdbdbar\ and \bsbsbar\ mixing.  The tree decay
amplitudes, being dominated by virtual $W$ exchange, remain
essentially unaffected by new physics. Thus, even in the presence of
new physics, the measured values of $|V_{cb}|$ and $|V_{ub}|$
correspond to their true SM values, so that two sides of the UT are
unaffected. However, the third side, which depends on $|V_{td}|$, will
in general be affected by new physics.  Furthermore, the measurements
of $\abseps$ and \bsbsbar\ mixing, which provide additional
constraints on the UT, will also be affected.  If Nature is kind, the
unitarity triangle, as constructed from direct measurements of
$\alpha$, $\beta$ and $\gamma$, will be inconsistent with that
obtained from independent measurements of the sides. If this were to
happen, it would be clear evidence for the presence of physics beyond
the SM, and would be most exciting. In such a case, the new physics is
also expected to modify the decay rates and distributions of rare
$B$-decays such as $B \to X_s \gamma$, $B \to X_s \ell^+ \ell^-$ and
$B \to X_s \nu \bar{\nu}$, and of related exclusive decays.
(Similarly, the corresponding decays dominated by the $b \to d$
transitions may also be affected.)

One type of new physics which has been extensively studied is
supersymmetry (SUSY). A hint suggesting that SUSY might indeed be
around the corner is the gauge-coupling unification: a supersymmetric
grand unified theory does better than its non-supersymmetric
counterpart. A great deal of effort has gone into a systematic study
of the pattern of flavour violation in SUSY, in particular in the
flavour-changing neutral-current processes in $K$ and $B$ decays. We
shall concentrate here on the minimal supersymmetric standard model
(MSSM), and update the anticipated profile of the UT and CP-phases
which we presented earlier\cite{AL99}. Of particular interest here is
the scenario of minimal supersymmetric flavour violation\cite{CDGG98},
which involves, in addition to the SM degrees of freedom, charged
Higgs bosons, a light stop (assumed right-handed) and a light
chargino, with all other degrees of freedom assumed heavy and hence
effectively integrated out.  This scenario can be embedded in
supergravity (SUGRA) models with gauge-mediated supersymmetry
breaking, in which the first two squark generations and the gluinos
are assumed heavy. Regardless of which variant is used, the key
assumption in our analysis is that there are no new phases in the
couplings -- although there are many new contributions to meson mixing
and rare decays, all are proportional to the same combination of CKM
matrix elements as found in the SM. As explained above, in this class
of models measurements of the CP phases will yield the true SM values
for these quantities. However, measurements of meson mixing and rare
decays will be affected by the presence of this new physics.

In Section 2, we discuss the profile of the unitarity triangle within
the SM. We describe the input data used in the fits and present the
allowed region in $\rho$--$\eta$ space, as well as the
presently-allowed ranges for the CP angles $\alpha$, $\beta$ and
$\gamma$. We turn to supersymmetric models in Section 3. We review
several variants of the MSSM, in which the new CP-violating phases are
essentially zero.  We also discuss the NLO corrections in such models
and show that the SUSY contributions to \kkbar, \bdbdbar\ and
\bsbsbar\ mixing are of the same form and can be characterized by a
single parameter $f$. We compare the profile of the unitarity triangle
in SUSY models, for various values of $f$, with that of the SM. We
conclude in Section 4.

\section{Unitarity Triangle: SM Profile}

\subsection{Input Data}

We briefly describe below the experimental and theoretical data which
constrain the CKM parameters. (For more details, we refer the reader
to\cite{AL99}.) A summary can be found in Table \ref{datatable}.

\begin{itemize}
  
\item The CKM parameters $\lambda$, $A$, $\rho$ and $\eta$ are
directly constrained through measurements of the CKM elements $\vert
V_{us} \vert = \lambda$\cite{PDG98}, $\absvcb$\cite{PDG98} and
$|V_{ub}/V_{cb}|$\cite{Parodiconf98}. In our fits we ignore the small
error on $\lambda$. Also, the error on $|V_{ub}/V_{cb}|$ includes some
theoretical model dependence.

\item {$ \abseps, \hat{B}_K$}: In the standard model, $\abseps$ is
essentially proportional to the imaginary part of the box diagram for
\kkbar\ mixing and is given by\cite{Burasetal}
\begin{eqnarray}
\abseps &=& \frac{G_F^2f_K^2M_KM_W^2}{6\sqrt{2}\pi^2\Delta M_K}
\hat{B}_K\left(A^2\lambda^6{\bar{\eta}}\right)
\bigl(y_c\left\{\hat{\eta}_{ct}f_3(y_c,y_t)-\hat{\eta}_{cc}\right\}
 \nonumber \\
&~& ~~~~~~~~~~~~~~+ 
~\hat{\eta}_{tt}y_tf_2(y_t)A^2\lambda^4(1-\bar{\rho})\bigr), 
\label{eps}
\end{eqnarray}
where $y_i\equiv m_i^2/M_W^2$, and the functions $f_2$ and $f_3$ can
be seen in\cite{InamiLim}.  Here, the $\hat{\eta}_i$ are QCD
correction factors, calculated at next-to-leading order in
Refs.\cite{HN94} ($\hat{\eta}_{cc}$),\cite{etaB} ($\hat{\eta}_{tt}$)
and\cite{HN95} ($\hat{\eta}_{ct}$).  The theoretical hadronic
uncertainty in the expression for $\abseps$ is in the
renormalization-scale independent parameter $\hat{B}_K$. In Table
\ref{datatable}, the $\abseps$ entry is taken from Ref.\cite{PDG98},
while that for $\hat{B}_K$ is based on lattice QCD methods, summarized
in Ref.\cite{Draper98}.

\item {$ \Delta M_d, \fbb$}: The mass difference $\Delta M_d$ is
calculated from the \bdbdbar\ box diagram, which is dominated by
$t$-quark exchange:
\beq
\label{bdmixing}
\Delta M_d = \frac{G_F^2}{6\pi^2}M_W^2M_B\left(\fbb\right)\hat{\eta}_B y_t
f_2(y_t) \vert V_{td}^*V_{tb}\vert^2~, \label{xd}
\eeq
where, using Eq.~(\ref{CKM}), $\vert V_{td}^*V_{tb}\vert^2=
A^2\lambda^{6} [\left(1-\bar{\rho}\right)^2+\bar{\eta}^2]$. Here,
$\hat{\eta}_B=0.55$ is the QCD correction, calculated in the
$\overline{MS}$ scheme\cite{etaB}. Consistency requires that the top
quark mass be rescaled from its pole (mass) value of $\mt =175 \pm 5$
GeV to the value $\overline{\mt}(\mt(pole))=165 \pm 5$ GeV in the
$\overline{MS}$ scheme. The slight dependence of $\hat{\eta}_B$ on
$\overline{\mt}(\mt(pole))$ in the range given here is ignored.  The
entry for $\Delta M_d$ in Table \ref{datatable} is taken from
Ref.\cite{Blaylock99}.

For the $B$ system, the hadronic uncertainty is given by $\fbb$,
analogous to $\hat{B}_K$ in the kaon system.  Present estimates of
this quantity are summarized in Ref.\cite{Draper98}, yielding $\fbd
\sqrt{\hat{B}_{B_d}} =(190 \pm 23)$ MeV in the quenched approximation.
The effect of unquenching is not yet understood completely. Taking the
MILC collaboration estimates of unquenching would increase the central
value of $\fbd \sqrt{\hat{B}_{B_d}}$ by $21$ MeV\cite{MILC98}. The
range of $\fbd \sqrt{\hat{B}_{B_d}}$ given in Table \ref{datatable} is
chosen to take all these considerations into account.

\item {$ \Delta M_s, \fbbs$}: Mixing in the \bsbsbar\ system is quite
similar to that in the \bdbdbar\ system. The \bsbsbar\ box diagram is
again dominated by $t$-quark exchange, and the mass difference between
the mass eigenstates $\delms$ is given by a formula analogous to that
of Eq.~(\ref{xd}):
\beq
\delms = \frac{G_F^2}{6\pi^2}M_W^2M_{B_s}\left(\fbbs\right)
\hat{\eta}_{B_s} y_t f_2(y_t) \vert V_{ts}^*V_{tb}\vert^2~.
\label{xs}
\eeq
Using the fact that $\vert V_{cb}\vert=\vert V_{ts}\vert$
(Eq.~(\ref{CKM})), it is clear that one of the sides of the unitarity
triangle, $\vert V_{td}/\lambda V_{cb}\vert$, can be obtained from the
ratio of $\delmd$ and $\delms$,
\beq
\frac{\delms}{\delmd} =
 \frac{\hat{\eta}_{B_s}M_{B_s}\left(\fbbs\right)}
{\hat{\eta}_{B_d}M_{B_d}\left(\fbb\right)}
\left\vert \frac{V_{ts}}{V_{td}} \right\vert^2.
\label{xratio}
\eeq
The only real uncertainty in this quantity is the ratio of hadronic
matrix elements $\fbbs/\fbb$. It is now widely accepted that the ratio
$\xi_s \equiv (f_{B_s} \sqrt{\hat{B}_{B_s}}) /
(f_{B_d}\sqrt{\hat{B}_{B_d}})$ is probably the most reliable of the
lattice-QCD estimates in $B$ physics. The value given Table
\ref{datatable} is based on Ref.\cite{Draper98}.

The present lower bound on $\Delta M_s$ is: $\Delta M_s > 14.3
~\mbox{(ps)}^{-1}$ (at $95\%$ C.L.)\cite{Blaylock99}. This bound has
been established using the so-called ``amplitude
method''\cite{Moser97}, and we follow this method in including the
current information about \bsbsbar\ mixing in the fits.

\end{itemize}

Referring to Table \ref{datatable}, we see that the quantities with
the largest errors are ${\hat\eta}_{cc}$ (28\%), ${\hat B}_K$ (16\%),
$|V_{ub}/V_{cb}|$ (15\%) and $\fbd\sqrt{\hat{B}_{B_d}}$ (19\%). Of
these, the latter three are extremely important in defining the
allowed $\rho$--$\eta$ region (the large error on ${\hat\eta}_{cc}$
does not affect the fit very much).  The errors on two of these
quantities --- ${\hat B}_K$ and $\fbd\sqrt{\hat{B}_{B_d}}$ --- are
purely theoretical in origin, and the error on $|V_{ub}/V_{cb}|$ has a
significant theoretical component (model dependence). Thus, the
present uncertainty in the shape of the unitarity triangle is due in
large part to theoretical errors.  Reducing these errors will be quite
important in getting a precise profile of the unitarity triangle and
the CP-violating phases.
%
\begin{table}[t]
\centering
\caption{ \it Data used in the CKM fits}
\vskip 0.1 in
\begin{tabular}{|c|c|} \hline
Parameter & Value \\
\hline
\hline
$\lambda$ & $0.2196$  \cr
$\vert V_{cb} \vert $ & $0.0395 \pm 0.0017$ \cr
$\vert V_{ub} / V_{cb} \vert$ & $0.093 \pm 0.014$ \cr
$\abseps$ & $(2.280 \pm 0.013) \times 10^{-3}$ \cr
$\Delta M_d$ & $(0.473 \pm 0.016)~(ps)^{-1}$ \cr
$\Delta M_s$ & $ > 14.3 ~(ps)^{-1}$ \cr 
$\overline{\mt}(\mt(pole))$ & $(165 \pm 5)$ GeV  \cr
$\overline{\mc}(\mc(pole))$ & $1.25 \pm 0.05$ GeV  \cr
$\hat{\eta}_B$ & $0.55$ \cr
$\hat{\eta}_{cc} $ & $1.38 \pm 0.53$  \cr
$\hat{\eta}_{ct} $ & $0.47 \pm 0.04$  \cr
$\hat{\eta}_{tt} $ & $0.57$  \cr
$\hat{B}_K$ & $0.94 \pm 0.15$  \cr
$\fbd\sqrt{\hat{B}_{B_d}} $ & $215 \pm 40$ MeV  \cr
$\xi_s $ & $1.14 \pm 0.06$   \cr
\hline
\end{tabular}
\label{datatable}
\end{table}

There are two other measurements which should be mentioned
here. First, the KTEV collaboration\cite{KTEV99} at Fermilab and the
NA48 collaboration\cite{NA48} at CERN have reported in 1999 new
measurements of direct CP violation in the $K$ sector through the
ratio $\epsilon^\prime/\epsilon$. Their results, together with those
of the earlier experiments NA31\cite{NA31} and E731\cite{E731} are as
follows:
\begin{eqnarray}
{\rm Re} (\epsilon^\prime/\epsilon) &=& ( 28.0 \pm 4.1) \times 10^{-4}
~~[\mbox{KTEV '99}],\\ \nonumber
&=& (18.5 \pm 7.3) \times 10^{-4}~~[\mbox{NA48 '99}],\\ \nonumber
& =& (23.0 \pm 6.5) \times 10^{-4} ~~[\mbox{NA31 '93}],\\ \nonumber
& =& (7.4 \pm 5.9) \times 10^{-4} ~~[\mbox{E731 '93}],
\end{eqnarray}
yielding the present world average\cite{herab-012} ${\rm Re}
(\epsilon^\prime/\epsilon)= (21.2 \pm 4.6) \times 10^{-4}$. This
combined result excludes the superweak model\cite{superweak}.

A great deal of theoretical effort has gone into calculating this
quantity at next-to-leading order accuracy in the
SM\cite{Buraseps}. However, numerical
estimates require a number of non-perturbative parameters, which are
at present poorly known\cite{BS98}, yielding an
theoretical uncertainty which is larger than an order of magnitude.
Thus, whereas $\epsilon^\prime/\epsilon$ represents a landmark
measurement, removing the superweak model of Wolfenstein from further
consideration, its impact on CKM phenomenology, particularly in
constraining the CKM parameters, is marginal as
$\epsilon^\prime/\epsilon$ is dominated by non-perturbative
uncertainties.

Second, the CDF collaboration has recently made a measurement of $\sin
2\beta$\cite{CDF99}. In the Wolfenstein parametrization, $-\beta$ is
the phase of the CKM matrix element $V_{td}$. From Eq.~(\ref{CKM}) one
can readily find that
\beq
\sin (2 \beta) = \frac{2\bar{\eta}(1-\bar{\rho})}{(1-\bar{\rho})^2 +
\bar{\eta}^2} ~.
\eeq
Thus, a measurement of $\sin 2\beta$ would put a strong contraint on
the parameters $\bar{\rho}$ and $\bar{\eta}$. However, the CDF
measurement gives\cite{CDF99}
\beq
\sin 2\beta = 0.79^{+0.41}_{-0.44} ~,
\eeq
or $\sin 2\beta > 0$ at 93\% C.L. As we will see in the next section,
this constraint is quite weak -- the indirect measurement (reported
here) already constrains $0.53 \le \sin 2\beta \le 0.93$ at the 95\%
C.L.\ in the SM.  In view of this, and given that it is not clear how
to combine the above measurement (which allows for unphysical values
of $\sin 2\beta$) with the other data, we have not included this
measurement in our fits.

\subsection{SM Fits}

In order to find the allowed region in $\bar{\rho}$--$\bar{\eta}$
space, i.e.\ the allowed shapes of the unitarity triangle, the
computer program MINUIT is used to fit the parameters to the
constraints described above. In the fit, we allow ten parameters to
vary: $\bar{\rho}$, $\bar{\eta}$, $A$, $m_t$, $m_c$, $\eta_{cc}$,
$\eta_{ct}$, $f_{B_d} \sqrt{\hat{B}_{B_d}}$, $\hat{B}_K$, and
$\xi_s$. The $\Delta M_s$ constraint is included using the amplitude
method.  The allowed (95\% C.L.) $\bar\rho$--$\bar\eta$ region is
shown in Fig.~\ref{rhoeta1}. The triangle drawn is to facilitate our
discussions, and corresponds to the central values of the fits,
$(\alpha,\beta,\gamma) = (93^\circ,24^\circ,63^\circ)$.

\begin{figure}
\vskip -1.0truein
\centerline{\epsfxsize 3.5 truein \epsfbox {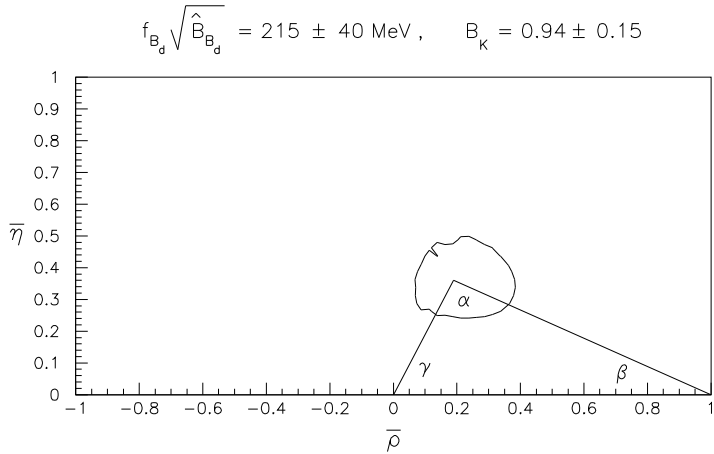}}
\vskip -1.5truein
\caption{Allowed region in $\bar\rho$--$\bar\eta$ space in the SM,
from a fit to the ten parameters discussed in the text and given in
Table \protect{\ref{datatable}}. The solid line represents the region
with $\chi^2=\chi_{min}^2+6$ corresponding to the 95\% C.L.\
region. The triangle shows the best fit.}
\label{rhoeta1}
\end{figure}

The CP angles $\alpha$, $\beta$ and $\gamma$ can be measured in
CP-violating rate asymmetries in $B$ decays.  These angles can be
expressed in terms of $\bar{\rho}$ and $\bar{\eta}$. Thus, different
shapes of the unitarity triangle are equivalent to different values of
the CP angles. Referring to Fig.~\ref{rhoeta1}, the allowed ranges at
95\% C.L. are given by
\beq
75^\circ \le \alpha \le 121^\circ ~~,~~~~
16^\circ \le \beta \le 34^\circ ~~,~~~~
38^\circ \le \gamma \le 81^\circ ~~,
\label{CPangleregion}
\eeq
or, equivalently,
\beq
-0.88 \le  \sin 2\alpha  \le 0.50 ~~,~~~~
0.53  \le  \sin 2\beta  \le 0.93  ~~,~~~~
0.38  \le  \sin^2 \gamma  \le 0.98 ~~.
\label{smabgrange}
\eeq

Of course, the values of $\alpha$, $\beta$ and $\gamma$ are
correlated, i.e.\ they are not all allowed simultaneously. After all,
the sum of these angles must equal $180^\circ$. We illustrate these
correlations in Figs.~\ref{alphabetacorrsm} and
\ref{alphagammacorrsm}. In both of these figures, the SM plot is
labelled by $f=0$. Fig.~\ref{alphabetacorrsm} shows the allowed region
in $\sin 2\alpha$--$\sin 2\beta$ space allowed by the data. And
Fig.~\ref{alphagammacorrsm} shows the allowed (correlated) values of
the CP angles $\alpha$ and $\gamma$. This correlation is roughly
linear, due to the relatively small allowed range of $\beta$
[Eq.~(\ref{CPangleregion})].

The allowed ranges for the CKM-parameters obtained from our unitarity
fits can be compared with those obtained by other groups. For example,
concentrating on $\sin 2 \alpha$ and $\sin 2 \beta$, Plaszczynski and
Schune get the following (95\% C.L.)  ranges\cite{PS99}:
\beq
-0.95 \le \sin 2\alpha \le 0.50 ~~,~~~~
0.50  \le \sin 2\beta \le 0.85 ~~,
\eeq
which are very similar to the ranges obtained by us for these
quantities [Eq.~(\ref{smabgrange})]. While there are smallish
differences in the input parameters from experimental measurements,
the real difference in the two fits lies in the incorporation of the
theoretical uncertainties.  We have treated theoretical and
experimental errors on the same footing. On the other hand,
Plaszczynski and Schune have scanned over a ``reasonable range" of
theoretical parameters, determined the allowed contours for fixed
values of these parameters and taken the envelope of all the
individual contours obtained in the allowed range. Of course, the size
of the resulting envelope depends on the assumed theoretical range, so
that a certain amount of subjectivity is already embedded. Given that
the parametric input in the present analysis and in\cite{PS99} are
similar, the closeness of the two fits implies that they do not depend
sensitively on the prescription for handling theoretical errors.

\begin{figure}
\vskip -2.0truein
\centerline{\epsfxsize 6.0 truein \epsfbox {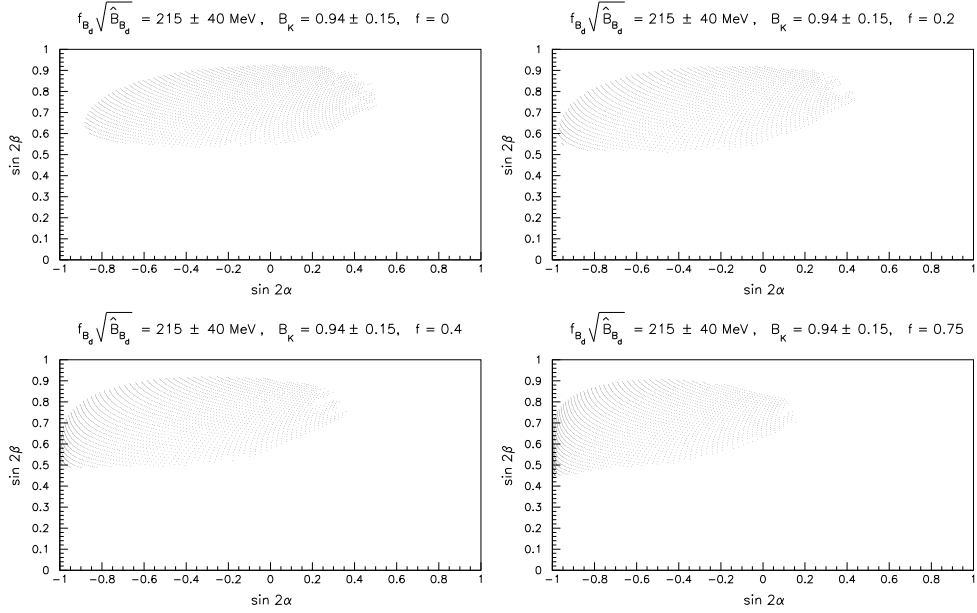}}
\vskip -3.7truein
\caption{Allowed 95\% C.L. region of the CP-violating quantities 
  $\sin 2\alpha$ and $\sin 2\beta$, from a fit to the data given in
  Table \protect{\ref{datatable}}. The upper left plot ($f=0$)
  corresponds to the SM, while the other plots ($f=0.2$, 0.4, 0.75)
  correspond to various SUSY models.}
\label{alphabetacorrsm}
\end{figure}

In fact, one can turn the argument around: with improved limits on (or
an actual measurement of) $\delms$, the theoretical errors on $\fbd
\sqrt{\hat{B}_{B_d}}$ and ${\hat B}_K$ can be effectively reduced. To
quantify these remarks, we examine the presently-allowed correlation
in the parameters $\hat{B}_K$ and $\fbd \sqrt{\hat{B}_{B_d}}$ which
follows from our fits in the SM. Recall that the theoretically-allowed
ranges for these quantities are $\fbd \sqrt{\hat{B}_{B_d}}= 215 \pm
40$ MeV and $\hat{B}_K= 0.94 \pm 0.15$. Rather than present the 95\%
c.l.\ region in the $\bar\rho$--$\bar\eta$ plane (Fig.~\ref{rhoeta1}),
we use the fits to find the allowed 95\% c.l.\ region in the
$\hat{B}_K$--$\fbd\sqrt{\hat{B}_{B_d}}$ plane. The results are shown in
Fig.~\ref{fbbkcorr}, where we have allowed the hadronic parameters to
vary in the range $135~\mbox{MeV}~\leq \fbd \sqrt{\hat{B}_{B_d}}\leq
295$ MeV and $0.64 \leq {\hat B}_K \leq 1.24$, which corresponds to
allowing a $\pm 2 \sigma$ uncertainty on each. (Note that there
appears to be some structure near the solid line on the left-hand side
of the figure. This is a numerical artifact due to the binning of the
$\delms$ data, and can be ignored. Only the solid line is important.)
Only values of $\fbd \sqrt{\hat{B}_{B_d}}$ and ${\hat B}_K$ which lie
between the two solid lines in Fig.~\ref{fbbkcorr} are allowed at the
95\% C.L. Note that present data do not allow a value $\fbd
\sqrt{\hat{B}_{B_d}} \leq 165$ MeV. Likewise, values of $\fbd
\sqrt{\hat{B}_{B_d}}$ in excess of $230$ MeV are highly correlated
with the value of ${\hat B}_K$. Thus, no values in excess of $230$ MeV
are allowed for $\fbd\sqrt{B_{B_d}}$ if $\hat{B}_K \leq 0.6$ in the
SM. This is very similar (though not identical) to the correlation
shown in Ref.\cite{PS99}.

\begin{figure}
\vskip -2.0truein
\centerline{\epsfxsize 6.0 truein \epsfbox {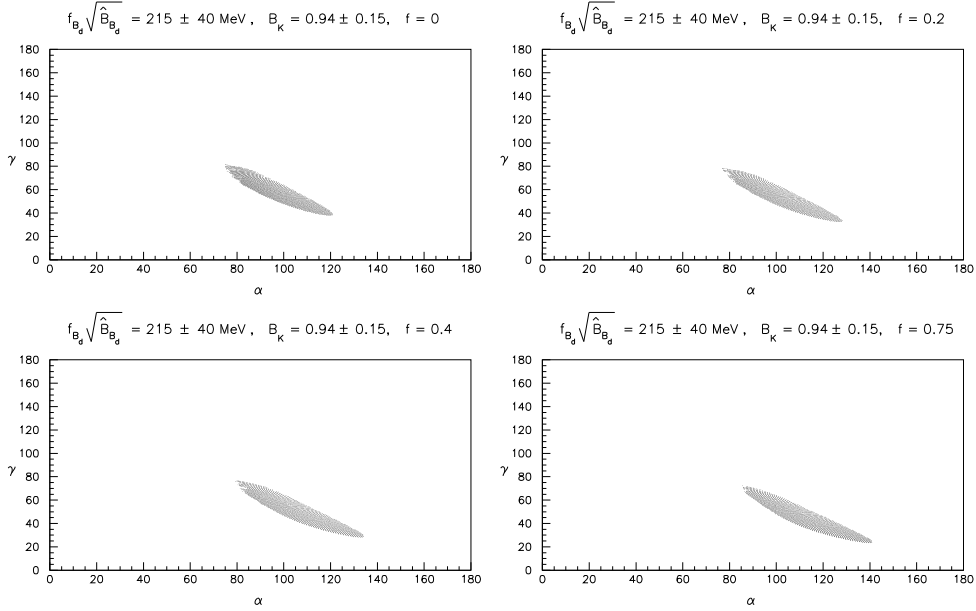}}
\vskip -3.7truein
\caption{Allowed 95\% C.L. region of the CP-violating quantities 
  $\alpha$ and $\gamma$, from a fit to the data given in Table
  \protect{\ref{datatable}}. The upper left plot ($f=0$) corresponds
  to the SM, while the other plots ($f=0.2$, 0.4, 0.75) correspond to
  various SUSY models.}
\label{alphagammacorrsm}
\end{figure}

\begin{figure}
\vskip -0.4truein
\centerline{\epsfxsize 2.5 truein \epsfbox {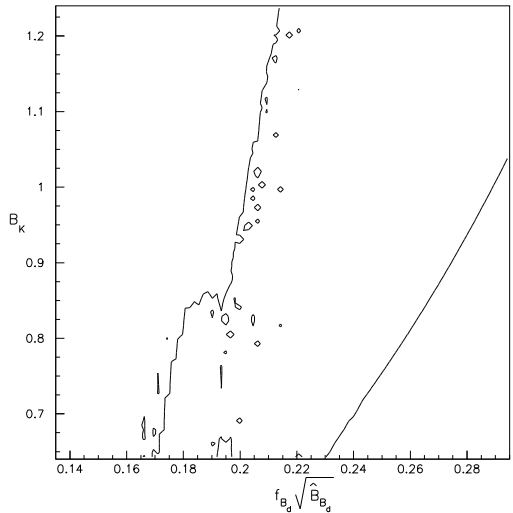}}
\vskip -0.5truein
\caption{Allowed 95\% C.L. region of the non-perturbative quantities
$\hat{B}_K$ and $\fbd \sqrt{\hat{B}_{B_d}}$ which results from the CKM
fits in the SM.}
\label{fbbkcorr}
\end{figure}

Of course, one obtains more stringent constraints on the CKM
parameters if significantly reduced errors are assumed for the input
parameters.  For example, a recent fit\cite{herab-012}, assuming
$\delta V_{ub}/V_{cb}= \pm 8.8\%$ (compared to $\delta
V_{ub}/V_{cb}=\pm 15\%$ used here), and $\fbd \sqrt{\hat{B}_{B_d}} =
220 \pm 28 ~\mbox{MeV}$ (as opposed to $\fbd \sqrt{\hat{B}_{B_d}} =
215 \pm 40 ~\mbox{MeV}$ in Table \ref{datatable}), leads to a more
precise determination of the apex of the unitarity triangle. In turn,
this yields the following 95\% C.L. ranges for the CP
asymmetries\cite{herab-012}:
\beq
-0.73 \le \sin 2\alpha \le 0.26 ~~,~~~~
0.63  \le \sin 2\beta \le 0.81  ~~,~~~~
0.51  \le \sin^2 \gamma \le 0.93 ~~.
\eeq

\section{Unitarity Triangle: A SUSY Profile}

In this section we update the profile of the unitarity triangle in
supersymmetric (SUSY) theories.  In general, minimal supersymmetric
standard models (MSSM) have three physical phases, apart from
the QCD vacuum parameter $\bar{\theta}_{\sss QCD}$ which we shall take
to be zero. The three phases are: (i) the CKM phase represented here
by the Wolfenstein parameter $\eta$, (ii) the phase $\theta_A=\arg
(A)$, and (iii) the phase $\theta_\mu=\arg (\mu)$\cite{DGH85}. The
last two phases, residing in the soft SUSY-breaking terms and in the
scalar superpotential, are
peculiar to SUSY models and their effects must be
taken into account in a general supersymmetric framework. In
particular, the CP-violating asymmetries which result from the
interference between mixing and decay amplitudes can produce
non-standard effects. Concentrating here on the $\Delta B=2$
amplitudes, two new phases $\theta_d$ and $\theta_s$ arise, which can
be parametrized as follows\cite{effsusy}:
\beq
\theta_{d,s} = \frac{1}{2} \arg \left(\frac{\langle B_{d,s} \vert {\cal 
H}_{eff}^{\sss SUSY} \vert \bar{B}_{d,s} \rangle}{\langle B_{d,s} \vert 
{\cal H}_{eff}^{\sss SM} \vert \bar{B}_{d,s} \rangle} \right) ~,
\eeq
where ${\cal H}^{\sss SUSY}$ is the effective Hamiltonian including
both the SM degrees of freedom and the SUSY contributions. Thus,
CP-violating asymmetries in $B$ decays would involve not only the
phases $\alpha$, $\beta$ and $\gamma$, defined previously, but
additionally $\theta_d$ or $\theta_s$. In other words, the SUSY
contributions to the real parts of $M_{12}(B_d)$ and $M_{12}(B_s)$ are
{\it no longer proportional} to the CKM matrix elements $V_{td}
V_{tb}^*$ and $V_{ts}V_{tb}^*$, respectively. If $\theta_d$ or
$\theta_s$ were unconstrained, one could not make firm predictions
about the CP asymmetries in SUSY models. In such a case, an analysis
of the profile of the unitarity triangle in such models would be
futile.

However, the experimental upper limits on the electric dipole moments
(EDMs) of the neutron and electron\cite{PDG98} do provide a constraint
on the phase $\theta_\mu$\cite{FOS95}. In supergravity (SUGRA) models
with {\it a priori} complex parameters $A$ and $\mu$, the phase
$\theta_\mu$ is strongly bounded with $\theta_\mu < 0.01
\pi$\cite{Nihei97}.

As for the phase $\theta_A$, it can be of $O(1)$ in the small
$\theta_\mu$ region, as far as the EDMs are concerned. However, in both the
$\Delta S=2$ and $\Delta B=2$ transitions, and for low-to-moderate
values of $\tan \upsilon$ \footnote{In supersymmetric jargon, the
quantity $\tan \beta$ is used to define the ratio of the two vacuum
expectation values (vevs) $\tan \beta \equiv v_u/v_d$, where
$v_d(v_u)$ is the vev of the Higgs field which couples exclusively to
down-type (up-type) quarks and leptons. (See, for example, the review
by Haber in Ref.\cite{PDG98}). However, in discussing flavour physics,
the symbol $\beta$ is traditionally reserved for one of the angles of
the unitarity triangle. To avoid confusion, we will call the ratio of
the vevs $\tan \upsilon$.}, it has been shown that $\theta_A$ does not
change the phase of either the matrix element $M_{12}(K)$\cite{DGH85}
or of $M_{12}(B)$\cite{Nihei97}. Hence, in SUGRA models, $\arg
M_{12}(B)|_{\sss SUGRA} = \arg M_{12}(B)|_{\sss SM}=\arg (\xi_t^2)$,
where $\xi_t=V_{td}^*V_{tb}$. Likewise, the phase of the SUSY
contribution in $M_{12}(K)$ is aligned with the phase of the
$t\bar{t}$-contribution in $M_{12}(K)$, given by $\arg
(V_{td}V_{ts}^*)$. 

Thus, in SUGRA models, one can effectively set $\theta_d \simeq 0$ and
$\theta_s \simeq 0$, so that the CP-violating asymmetries give
information about the SM phases $\alpha$, $\beta$ and $\gamma$. Hence,
an analysis of the UT and CP-violating phases $\alpha$, $\beta$ and
$\gamma$ can be carried out in a very similar fashion as in the SM,
taking into account the additional contributions to $M_{12}(K)$ and
$M_{12}(B)$.

\subsection{NLO Corrections to $\delmd$, $\delms$ and $\epsilon$ in Minimal 
SUSY Flavour Violation}

A number of SUSY models share the features mentioned in the previous
subsection, and the supersymmetric contributions to the mass
differences $M_{12}(B)$ and $M_{12}(K)$ have been analyzed in a number
of papers\cite{Nihei97,Brancoetal,Gotoetal96,Gotoetal97,Gotoetal98-1},
following the pioneering work of Ref.\cite{BBMR91}.  The SUSY
contributions to $\delmd$, $\delms$ and $\abseps$ in supersymmetric
theories can be incorporated in a simple form\cite{AL99}:
\begin{eqnarray}
\delmd &=& \delmd (SM) [ 1 +
f_d(m_{\chi_2^\pm},m_{\tilde{t}_R},
m_{H^\pm}, \tan \upsilon) ], \nonumber \\
\delms &=& \delms (SM) [ 1 +
f_s(m_{\chi_2^\pm},m_{\tilde{t}_R},
m_{H^\pm}, \tan \upsilon) ], \nonumber \\
\abseps &=& \frac{G_F^2f_K^2M_KM_W^2}{6\sqrt{2}\pi^2\Delta M_K}
\hat{B}_K\left(A^2\lambda^6\bar{\eta}\right)
\bigl(y_c\left\{\hat{\eta}_{ct}f_3(y_c,y_t)-\hat{\eta}_{cc}\right\}
 \nonumber \\
&~& +
~\hat{\eta}_{tt}y_tf_2(y_t)[1 + f_\epsilon
(m_{\chi_2^\pm},m_{\tilde{t}_2}, m_{H^\pm},
\tan \upsilon)] A^2\lambda^4(1-\bar{\rho})\bigr).
\label{susyformel}
\end{eqnarray}
The quantities $f_d$, $f_s$ and $f_\epsilon$ can be expressed as
\beq
\label{fis}
f_d = f_s= \frac{\hat{\eta}_{2,S}(B)}{\hat{\eta}_B} R_{\Delta_d}(S) ~,~~~~
f_\epsilon = \frac{\hat{\eta}_{2,S}(K)}{\hat{\eta}_{tt}}R_{\Delta_d}(S) ~,
\eeq
where $R_{\Delta_d}(S)$ is defined as
\beq
R_{\Delta_d}(S) \equiv { \Delta M_d(SUSY) \over \Delta M_d(SM) } (LO)
= {S \over y_t f_2(y_t)} ~.
\eeq
The supersymmetric function $S$ is given in Ref.\cite{BBMR91},
and the NLO functions $\hat{\eta}_{2,S}(B)$ and $\hat{\eta}_{2,S}(K)$
can be found in Ref.\cite{KS98}. The
functions $f_{i}$, $i=d,s,\epsilon$ are all positive definite, i.e.\
the supersymmetric contributions add {\it constructively} to the SM
contributions in the entire allowed supersymmetric parameter space.
The two QCD correction factors appearing in Eq.~(\ref{fis}) are
numerically very close to one another, with
$\hat{\eta}_{2,S}(B)/\hat{\eta}_B \simeq
{\hat\eta}_{2,S}(K)/\hat{\eta}_{tt} = 0.93$\cite{AL99}. Thus, to an
excellent approximation, one has $f_d = f_s = f_\epsilon \equiv f$.

How big can $f$ be? This quantity is a function of the masses of the
top squark, chargino and the charged Higgs, $m_{\tilde{t}_R}$,
$m_{\tilde{\chi}^\pm_2}$ and $m_{H^\pm}$, respectively, as well as of
$\tan\upsilon$. The maximum allowed value of $f$ depends on the model
(minimal SUGRA, non-minimal SUGRA, MSSM with constraints from EDMs,
etc.). From the published results we conclude that typically $f$ can
be as large as $0.45$ in non-minimal SUGRA models for low $\tan
\upsilon$ (typically $\tan \upsilon=2$)\cite{Gotoetal98-1}, and
approximately half of this value in minimal SUGRA
models\cite{Nihei97,Gotoetal97,Gotoetal98-1}. Relaxing the SUGRA mass
constraints, admitting complex values of $A$ and $\mu$ but
incorporating the EDM constraints, and imposing the constraints
mentioned above, $f$ could be larger\cite{BK98-1}. In all cases, the
value of $f$ decreases with increasing $\tan \upsilon$ or increasing
$m_{\tilde{\chi}^\pm_2}$ and $m_{\tilde{t}_R}$, as noted above.

\subsection{SUSY Fits}

For the SUSY fits, we use the same program as for the SM fits, except
that the theoretical expressions for $\Delta M_d$, $\Delta M_s$ and
$\abseps$ are modified as in Eq.~(\ref{susyformel}). We compare the
fits for four representative values of the SUSY function $f$ --- 0,
0.2, 0.4 and 0.75 --- which are typical of the SM, minimal SUGRA
models, non-minimal SUGRA models, and non-SUGRA models with EDM
constraints, respectively.

\begin{figure}
\vskip -1.0truein
\centerline{\epsfxsize 3.5 truein \epsfbox {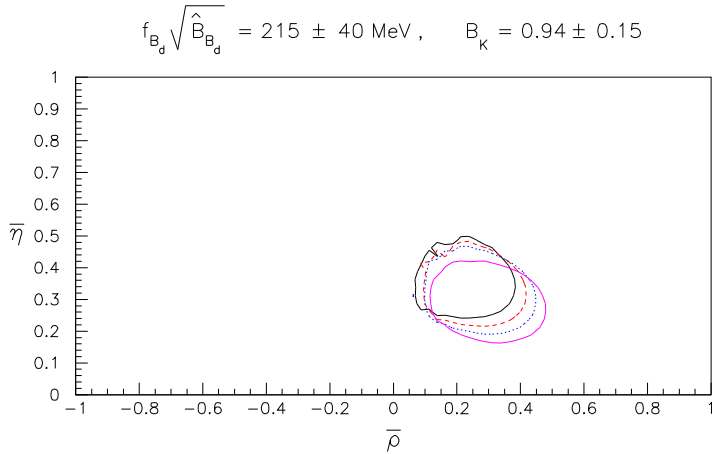}}
\vskip -1.5truein
\caption{Allowed 95\% C.L. region in $\rho$--$\eta$ space in the SM 
  and in SUSY models, from a fit to the data given in Table
  \protect{\ref{datatable}}. From left to right, the allowed regions
  correspond to $f=0$ (SM, solid line), $f=0.2$ (long dashed line),
  $f=0.4$ (short dashed line), $f=0.75$ (dotted line).}
\label{sugratot}
\end{figure}

The allowed 95\% C.L. regions for these four values of $f$ are all
plotted in Fig.~\ref{sugratot}. As is clear from this figure, there is
still a considerable overlap between the $f=0$ (SM) and $f=0.75$
regions. However, there are also regions allowed for one value of $f$
which are excluded for another value. Thus a sufficiently precise
determination of the unitarity triangle might be able to exclude
certain values of $f$ (including the SM, $f=0$).

\begin{table}
\hfil
\caption{Allowed 95\% C.L. ranges for the CP phases $\alpha$, $\beta$
and $\gamma$, as well as their central values, from the CKM fits in
the SM $(f=0)$ and supersymmetric theories, characterized by the
parameter $f$ defined in the text.}
\vbox{\offinterlineskip
\halign{&\vrule#&
 \strut\quad#\hfil\quad\cr
\noalign{\hrule}
height2pt&\omit&&\omit&&\omit&&\omit&&\omit&\cr 
& $f$ && $\alpha$ && $\beta$ && $\gamma$ && $(\alpha,\beta,\gamma)_{\rm cent}$ & \cr
height2pt&\omit&&\omit&&\omit&&\omit&&\omit&\cr 
\noalign{\hrule}
height2pt&\omit&&\omit&&\omit&&\omit&&\omit&\cr
& $f=0$ (SM) && $75^\circ$ -- $121^\circ$ && $16^\circ$ -- $34^\circ$ &&
$38^\circ$ -- $81^\circ$ && $(93^\circ, 24^\circ, 63^\circ)$ & \cr
& $f=0.2$ && $77^\circ$ -- $128^\circ$ && $15^\circ$ -- $33^\circ$ &&
$32^\circ$ -- $78^\circ$ && $(102^\circ, 24^\circ, 54^\circ)$ & \cr
& $f=0.4$ && $79^\circ$ -- $134^\circ$ && $15^\circ$ -- $33^\circ$ &&
$28^\circ$ -- $77^\circ$ && $(108^\circ, 23^\circ, 49^\circ)$ & \cr
& $f=0.75$ && $86^\circ$ -- $141^\circ$ && $13^\circ$ -- $33^\circ$ &&
$23^\circ$ -- $72^\circ$ && $(120^\circ, 21^\circ, 39^\circ)$ & \cr
height2pt&\omit&&\omit&&\omit&&\omit&&\omit&\cr 
\noalign{\hrule}}}
\label{cpasym1}
\end{table}

{}From Fig.~\ref{sugratot} it is clear that a measurement of the CP
angle $\beta$ will {\it not} distinguish among the various values of
$f$. Rather, it is the measurement of $\gamma$ or $\alpha$ which has
the potential to rule out certain values of $f$. As $f$ increases, the
allowed region moves slightly down and towards the right in the
$\bar{\rho}$--$\bar{\eta}$ plane, corresponding to smaller values of
$\gamma$ (or equivalently, larger values of $\alpha$). We illustrate
this in Table~\ref{cpasym1}, where we present the allowed ranges of
$\alpha$, $\beta$ and $\gamma$, as well as their central values
(corresponding to the preferred values of $\bar{\rho}$ and
$\bar{\eta}$), for each of the four values of $f$. From this Table, we
see that the allowed range of $\beta$ is largely insensitive to the
model. Conversely, the allowed values of $\alpha$ and $\gamma$ do
depend somewhat strongly on the chosen value of $f$. Note, however,
that one is not guaranteed to be able to distinguish among the various
models: as mentioned above, there is still significant overlap among
all four models. Thus, depending on what values of $\alpha$ and
$\gamma$ are obtained, we may or may not be able to rule out certain
values of $f$.

\begin{table}
\caption{Allowed 95\% C.L. ranges for the CP asymmetries $\sin
2\alpha$, $\sin 2\beta$ and $\sin^2 \gamma$, from the CKM fits in the
SM $(f=0)$ and supersymmetric theories, characterized by the parameter
$f$ defined in the text.}  
\hfil \vbox{\offinterlineskip
\halign{&\vrule#& \strut\quad#\hfil\quad\cr \noalign{\hrule}
height2pt&\omit&&\omit&&\omit&&\omit&\cr & $f$ && $\sin 2\alpha$ &&
$\sin 2\beta$ && $\sin^2 \gamma$ & \cr
height2pt&\omit&&\omit&&\omit&&\omit&\cr \noalign{\hrule}
height2pt&\omit&&\omit&&\omit&&\omit&\cr & $f=0$ (SM) && $-$0.88 --
0.50 && 0.53 -- 0.93 && 0.38 -- 0.98 & \cr & $f=0.2$ && $-$0.97 --
0.44 && 0.51 -- 0.92 && 0.29 -- 0.96 & \cr & $f=0.4$ && $-$1.00 --
0.36 && 0.49 -- 0.92 && 0.22 -- 0.95 & \cr & $f=0.75$ && $-$1.00 --
0.16 && 0.44 -- 0.91 && 0.16 -- 0.91 & \cr
height2pt&\omit&&\omit&&\omit&&\omit&\cr \noalign{\hrule}}}
\label{cpasym2}
\end{table}

For completeness, in Table~\ref{cpasym2} we present the corresponding
allowed ranges for the CP asymmetries $\sin 2\alpha$, $\sin 2\beta$
and $\sin^2 \gamma$. Again, we see that the allowed range of $\sin
2\beta$ is largely independent of the value of $f$. On the other hand,
as $f$ increases, the allowed values of $\sin 2\alpha$ become
increasingly negative, while those of $\sin^2 \gamma$ become smaller.
 
The allowed (correlated) values of the CP angles for various values of
$f$ can be clearly seen in Figs.~\ref{alphabetacorrsm} and
\ref{alphagammacorrsm}. As $f$ increases from 0 (SM) to 0.75, the
change in the allowed $\sin 2\alpha$--$\sin 2\beta$
(Fig.~\ref{alphabetacorrsm}) and $\alpha$--$\gamma$
(Fig.~\ref{alphagammacorrsm}) regions is quite significant.

\section{Conclusions}

In the very near future, CP-violating asymmetries in $B$ decays will
be measured at $B$-factories, HERA-B and hadron colliders. Such
measurements will give us crucial information about the interior
angles $\alpha$, $\beta$ and $\gamma$ of the unitarity triangle. If we
are lucky, there will be an inconsistency in the independent
measurements of the sides and angles of this triangle, thereby
revealing the presence of new physics.

An interesting possibility, from the point of view of making
predictions, are models which contribute to $B^0$--${\overline{B^0}}$
mixings and $\abseps$, but without new phases. One type of new physics
which does just this is supersymmetry (SUSY). There are some SUSY
models which do contain new phases, but they suffer from a lack of
predictivity. However, there is also a large class of SUSY models with
no new phases.  In these models, there are new, supersymmetric
contributions to \kkbar, \bdbdbar\ and \bsbsbar\ mixing. The key
ingredient in our analysis is the fact that these contributions, which
add constructively to the SM, depend on the SUSY parameters in
essentially the same way. That is, so far as an analysis of the
unitarity triangle is concerned, there is a single parameter, $f$,
which characterizes the various SUSY models within this class of
models ($f=0$ corresponds to the SM).

We have therefore updated the profile of the unitarity triangle in
both the SM and some variants of the MSSM. We have used the latest
experimental data on $|V_{cb}|$, $|V_{ub}/V_{cb}|$, $\Delta M_d$ and
$\Delta M_s$, as well as the latest theoretical estimates (including
errors) of $\hat{B}_K$, $\fbd\sqrt{\hat{B}_{B_d}}$ and $\xi_s \equiv
\fbd\sqrt{\hat{B}_{B_d}}/\fbs\sqrt{\hat{B}_{B_s}}$. In addition to
$f=0$ (SM), we considered three SUSY values of $f$: 0.2, 0.4 and
0.75, representing the minimal SUGRA models, non-minimal SUGRA
models, and non-SUGRA models with EDM constraints, respectively.

We first considered the profile of the unitarity triangle in the SM,
shown in Fig.~\ref{rhoeta1}. We then compared the SM with the
different SUSY models. The result can be seen in
Fig.~\ref{sugratot}. As $f$ increases, the allowed region moves
slightly down and to the right in the $\bar\rho$--$\bar\eta$
plane. The main conclusion from this analysis is that the measurement
of the CP angle $\beta$ will not distinguish among the SM and the
various SUSY models -- the allowed region of $\beta$ is virtually the
same in all these models. On the other hand, the allowed ranges of
$\alpha$ and $\gamma$ do depend on the choice of $f$. For example,
larger values of $f$ tend to favour smaller values of $\gamma$. The
dependence of the CP angles on the value of $f$ is illustrated clearly
in Tables \ref{cpasym1} and \ref{cpasym2}. Thus, with measurements of
$\gamma$ or $\alpha$, we may be able to rule out certain values of $f$
(including the SM, $f=0$). However, we also note that there is no
guarantee of this happening -- at present there is still a significant
region of overlap among all four models.

\bigskip
\noindent
{\bf Acknowledgements}:
\bigskip

One of us (A.A.) would like to thank the organizers of the DAPHNE '99
workshop, in particular Giorgio Capon and Gino Isidori, for 
their kind hospitality. The work of D.L. was financially supported by
NSERC of Canada.
  

\end{document}